\documentclass[12pt,preprint]{aastex}

\usepackage{amsfonts}
\usepackage{amsmath}
\usepackage{amssymb}
\usepackage{pdflscape}
\usepackage{rotating}
\usepackage{natbib}
\usepackage{txfonts}
\usepackage{lineno}
\usepackage{longtable}
\usepackage[small,bf]{subfigure}
\usepackage{multirow}
\usepackage{xspace}
\usepackage[pdftitle={Cyg-X3-neutrinos-V24}, colorlinks=true, citecolor=blue, linkcolor=blue, urlcolor=blue]{hyperref}
\newcommand{\de}{\mathrm{d}}

\renewcommand{\rho}{\varrho}

\def \gray {$\gamma$-ray\xspace}
\def \grays {$\gamma$-rays\xspace}
\def \flx {$\rm{photons~cm^{-2}~s^{-1}}$\xspace}
\def \lat {\textit{Fermi}-LAT\xspace}
\def \fermi {\textit{Fermi}\xspace}
\def \cyg {\mbox{Cyg X-3}\xspace}

\shorttitle{Hadronic HE $\gamma$-ray and $\nu$ emission from Cygnus X-3}
\shortauthors{N. Sahakyan et al.}

\begin{document}

\title{Hadronic Gamma-Ray and Neutrino Emission from Cygnus X-3}

\author{N. Sahakyan\altaffilmark{1,2}}
\affil{$^1$ ICRANet-Yerevan, Marshall Baghramian Avenue, 24, Yerevan 0019, Republic of Armenia}
\affil{$^2$ ICRANet, Piazza della Repubblica 10, I-65122 Pescara, Italy}
\email{narek@icra.it}
\author{G. Piano\altaffilmark{3,4,5}}
\and
\author{M. Tavani\altaffilmark{3,4,5,6}}
\affil{$^3$ INAF/IAPS, via del Fosso del Cavaliere 100, I-00133 Roma, Italy}
\affil{$^4$ CIFS-Torino, viale Settimio Severo 3, I-10133 Torino, Italy}
\affil{$^5$ INFN-Roma ``Tor Vergata'', via della Ricerca Scientifica 1, I-00133 Roma, Italy}
\affil{$^6$ Dipartimento di Fisica, Universit\`a di Roma ``Tor Vergata'', via della Ricerca Scientifica 1,I-00133 Roma, Italy}

\begin{abstract}

Cygnus X-3 (Cyg X-3) is a remarkable Galactic microquasar
(X-ray binary) emitting from radio to $\gamma$-ray energies.
In this paper, we consider hadronic model of emission of
$\gamma$-rays above 100 MeV and their implications. We focus here on
the joint $\gamma$-ray and neutrino production resulting from
proton-proton interactions within the binary system.
We find that the required proton injection
kinetic power, necessary to explain the $\gamma$-ray flux
observed by AGILE and \textit{Fermi}-LAT, is $L_p \sim 10^{38}\:\rm{erg\:s^{-1}}$, a value
in agreement with the average bolometric luminosity of the hypersoft state
(when Cygnus X-3 was repeatedly observed to produce transient $\gamma$-ray activity).
If we assume an increase of the wind density at the superior conjunction, the
asymmetric production of $\gamma$-rays along the orbit
can reproduce the observed modulation. According to
observational constraints and our modelling, a maximal  flux of
high-energy neutrinos would be produced for an initial proton
distribution with a power-law index
$\alpha=2.4$. The predicted neutrino flux is almost two orders of
magnitude less than the 2-month IceCube sensitivity at $\sim$1
TeV.
If the protons are accelerated up to PeV energies, the
predicted neutrino flux for a prolonged ``soft X-ray state'' would be a factor of about 3 lower
than the 1-year IceCube sensitivity at $\sim$10 TeV. This study shows that,
for a prolonged soft state (as observed in 2006) possibly related with $\gamma$-ray activity
and a hard distribution of injected protons, Cygnus X-3 might be close to being detectable by
cubic-kilometer neutrino telescopes such as IceCube.
\end{abstract}


\section{Introduction}

\cyg was discovered in 1966 \citep{giacconi_67} as a bright X-ray source.
It is a high-mass X-ray binary (HMXB) located at an estimated distance of about 7-10 kpc \citep{bonnet_88,ling_09}.
The donor star is known to be a Wolf-Rayet (WR) star \citep{vankerk_92}
with a strong helium stellar wind \citep{szo_zdzia_08}, while it's still unclear whether
the accreting object is a neutron star or a stellar black hole \citep{vilhu_09},
even though a black hole scenario is favored \citep{szo_zdzia_08, szostek_08}.
The orbital period -- detected in infrared \citep{becklin_73}, X-ray
\citep{parsignault_72}, and \gray \citep{abdo_09} bands -- is very short (4.8 hours),
indicating that the compact object is completely enshrouded in the wind of the companion star
(orbital distance, $d \approx 3 \times 10^{11}$ cm).
\cyg is known to produce giant radio outburst (``major radio flares'') up to a few tens of Jy.
During these huge flares, milliarcsec-scale observations at cm wavelengths found an
expanding one-sided relativistic jet ($v \sim 0.81c$,), with an inclination to the line-of-sight
of $\lesssim14^{\circ}$ \citep{mioduszewski_01, tudose_10}.
High-energy \grays (HE \grays: $>$ 100 MeV) from \cyg have been
firmly detected by the new generation of space telescopes.
AGILE found evidence of transient \gray activity from \cyg
in 2009 \citep{tavani_09} as confirmed by the \lat
detections \citep{abdo_09}. Furthermore,  \lat
could determine also the orbital modulation of the \gray
emission \citep{abdo_09}. The photon spectrum (between 100 MeV
and 3 GeV) detected by AGILE during the peak flaring activity is
well-fitted by a power law with a photon index $\Gamma=2.0 \pm
0.2$. On the other hand, the average spectrum above 100 MeV
measured by \lat for two active windows (of about two months each)
gives $\Gamma=2.70 \pm 0.25$.
This difference could indicate a fast spectral hardening
of the high-energy emission during the short and intense \gray events
(lasting $\sim$1-2 days) detected by AGILE.
Nevertheless, the peak \gray luminosity detected above 100 MeV by
both AGILE and \lat corresponds to $L_{\gamma}\approx10^{36}~\rm{erg\:s^{-1}}$.
These observations provide direct evidence that extreme particle acceleration
occurs in \cyg in a transient fashion, most likely
associated with the relativistic jet ejection and/or propagation.
Both AGILE \citep{tavani_09, bulgarelli_12, piano_12} and
Fermi \citep{abdo_09, corbel_12} found the same multi-frequency
conditions for the \gray activity. In particular, \citet{piano_12}
found that the transient \gray emission detected by AGILE is
associated with very faint hard X-ray activity\footnote{The \gray
activity has been detected -- always during soft X-ray spectral
states -- when the 15-50 keV count rate detected by
\textit{Swift}/BAT (Burst Alert Telescope) was lower than
$0.02~\mathrm{counts}$ $\mathrm{cm^{-2}~s^{-1}}$ $(\approx0.091~\mathrm{Crab})$.} and generally
occurs a few days before intense major radio flares. The \gray
transient emission is observed when the system is moving into or
out of the quenched state, a characteristic state of \cyg that
generally precedes a major radio flare. The quenched state, which
has been found to be a key trigger condition for the \gray
activity, is characterized by a very low or undetectable level of
radio flux density and a bright soft X-ray emission with a
particular X-ray spectrum (the hypersoft spectrum,
\citealp{koljonen_10}).

At very high energies (VHE \grays: $>$ 250 GeV),
\cyg was observed by the MAGIC telescope for about 70 hr
between 2006, March and 2009, August. These observations
correspond to different X-ray/radio spectral states,
and also show periods of enhanced \gray emission
\citep{aleksic_10}. No TeV \grays from \cyg have been detected
and upper limits on the integrated \gray flux above 250 GeV
are $2.2 \times 10^{-12}$ \flx \citep{aleksic_10}. We notice
that the lack of evidence for detectable TeV emission from \cyg
can be due to the strong absorption of
these photons (through $\gamma\gamma$ absorption on ultraviolet
(UV) stellar photons from the WR star) or due to limited time of
observation. If \gray emission from \cyg is caused by
hadronic interactions, our knowledge about the source would be greatly improved by
the detection of HE neutrinos.
For proton-proton ($pp$) interactions, the emitted photons
and neutrinos can have comparable intensities. In
this case, high-energy photons can be absorbed and neutrinos
can freely escape from the source.
Neutrinos carry invaluable  information about the
existence (or absence) of energetic protons and shed light on
the location of the \gray production region.

Recently, the IceCube collaboration reported on
searches for neutrino sources at energies above 200 GeV in the
Northern sky of the Galactic plane (including \cyg), using the
data collected by the South Pole neutrino telescope, IceCube, and
AMANDA \citep{abbas_13}. Interestingly, it turns out that
during this period \cyg was observed both close to \gray flaring activity as well as in
different X-ray/radio states. A maximum likelihood test using a
time-dependent version of the unbinned likelihood ratio method
was applied to the IceCube data. As a result, no
evidence for a signal was found in the neutrino sample. The
90\% confidence level upper limits  on the differential muon
neutrino flux from \cyg for $E^{-2}$ and $E^{-3}$ spectra are:
$\rm{dN/dE} \leq 0.7 \times 10^{-11}~\rm{TeV^{-1} cm^{-2} s^{-1}}$
and $\rm{dN/dE} \leq 5 \times 10^{-11}~\rm{TeV^{-1} cm^{-2}
s^{-1}}$, respectively \citep{abbas_13}.

In this paper we focus on the hadronic scenario for high
energy emission from the microquasar \cyg. We assume that the jet
of \cyg accelerates both leptons and hadrons to high energies. The
accelerated protons escape from the jet and, interacting with
hadronic matter of the WR star, produce \grays and neutrinos. By
normalizing our model with the \gray emission of \cyg at its peak (using AGILE data)
and considering the MAGIC upper limits (that we interpret in terms of a strong $\gamma \gamma$
absorption on the stellar photons), the corresponding flux of HE
neutrinos is calculated and compared with the IceCube sensitivity.

This paper is organized as follows. In
section~\ref{sec2} the model for production of hadronic \grays is described:
in section~\ref{sec3} the spectra of \grays and neutrinos
produced in $pp$ interactions are calculated; the results are presented in section~\ref{sec4}.
The discussions and conclusions are presented in sections \ref{sec5} and \ref{sec6}.

\section{The model}\label{sec2}

The origin of HE \grays from microquasars can be
interpreted within both leptonic and hadronic scenarios. The jet
in the microquasars is a powerful particle accelerator
(electrons and/or protons), and the photon field and/or wind from
companion star can be a target for \gray production. In the leptonic scenario, HE
\grays are produced from inverse Compton (IC) scattering of
soft seed photons (from the companion star and from the accretion
disk) by energetic electrons. In the case of \cyg, the leptonic
scenario has been extensively discussed in the literature
\citep{dubus_10, zdziarski_12, piano_12}. In particular,
\citealp{piano_12} applied this picture to the AGILE observations
of \cyg in flaring states, showing how a leptonic scenario can
explain the spectral shape at GeV energies as well as the hard
X-ray emission at $\sim$100 keV observed during the transient
\gray activity.

By considering a hadronic picture, since
protons are characterized by a longer cooling time than electrons,
we can assume that protons are accelerated well above 10 TeV.
The detection of a periodic TeV \gray signal (which
could be evidence for production of TeV photons in a binary system)
would provide additional information on the problem of distinguishing
leptonic and hadronic contributions. If accelerated in the presence of a strong
cooling photon background, electrons would possibly produce VHE \grays by IC scattering.
On the other hand, hadrons would produce VHE \grays by interaction of a suitable gaseous target.
After escaping from the jet, protons can interact both with the X-ray photon field
from the accretion disk ($p\gamma$ interaction) as well as
with the hadronic component of the stellar wind ($pp$ interaction). In both cases,
a significant flux of TeV \grays and neutrinos are predicted:
while the \grays are absorbed (depending on the energy),
the neutrinos escape from the region with negligible absorption.
The emerging flux of HE neutrinos can significantly exceed
the \gray observed flux (approximately $\exp(\tau)$ times higher,
where $\tau$ is the optical depth of $\gamma\gamma$ absorption).
For example in the case of the microquasar LS 5039, the flux of HE neutrinos
can be as large as $1.6\times10^{-11}\:\rm{cm^{-2}\:s^{-1}}$ (for
energy greater than 1 TeV), above the sensitivity threshold of
experiments in the Mediterranean Sea \citep{ahar01}. Therefore the
binary systems could have a high flux of HE neutrinos,
which can be detected by the current generation of detectors.

Hadronic models based on $p\gamma$ interactions require
acceleration of protons in the inner jet up to the energies of
$10^{16}$ eV in order to produce a significant flux of HE \grays
and neutrinos. On the other hand, models based on $pp$
interactions can explain the observed \gray flux by requiring
lower energies of accelerated protons. Production of HE neutrinos
from $p\gamma$ interaction in \cyg have been discussed in
\citet{baerwald_12}: the expected number of neutrinos -- assuming
the IceCube sensitivity -- during 61 days is 0.02, corresponding
to a non-detectable flux.
An alternative hadronic scenario for production of \grays from \cyg
was discussed in \citealp{piano_12}, within the jet-wind interaction model.
They found that the constraints on the energetics of the system
are physically reasonable: the required jet kinetic power
is lower than the Eddington accretion limit for the source, and
the resulting spectral shape is consistent with the observed spectrum above 100 MeV.
In our model we assume that the energy budget in the jet is dominated by
the kinetic energy of an \textit{e-p} plasma and it contains a significant population
of protons. These particles are accelerated along the jet propagation (e.g via shock
acceleration) and can reach very high energies; in the binary frame this energy is even
higher (multiplied by the Lorentz factor of the jet). Due to the slowly cooling of the
protons in the jet, their maximum energy will be likely limited by the size of acceleration
region: the Larmor radius ($r_{L}$) of the protons should be contained in
the acceleration region $r_{L}\leq R_{jet}$ where $R_{jet}$ is the jet radius and $r_{L}=E_m/eB$, where $e$ is the elementary charge.
Depending on the efficiency of acceleration, the magnetic field ($B$), and the jet radius
($R_{jet}$), the proton maximum energy ($E_m$) can be as large as 100 TeV. However, the
accelerated protons can escape from the jet at some distance from the compact object
since the magnetic field gets weaker. In case this occurs in a binary system, the protons interact
with the dense wind of the WR star, producing neutral and charged pions via inelastic hadronic scattering.
The neutral pions subsequently decay in \grays, while muon and electron
neutrinos ($\nu_\mu, \nu_e$) are produced by the decays of
$\pi^\pm$ (e.g., $\pi^+ \rightarrow \mu^+ + \nu_\mu \rightarrow e^+ + \nu_e + \nu_\mu +
{\bar \nu}_\mu$). If we define $L_p$ as the luminosity of relativistic protons,
the corresponding luminosity of \grays is $L_{\gamma}\approx c_{pp}\:L_{p}$
where $c_{pp}$ is the energy transfer efficiency from relativistic protons
to secondary particles (for simplicity it is assumed that the escape time
of protons from the binary system is longer than the cooling time).
Similarly, expressing the acceleration power of protons in terms of the total jet power,
$L_{p}\approx \zeta\:L_{jet}$, one finds the following relation between jet power
and \gray luminosity, $L_{\gamma}\approx \zeta c_{pp}\:L_{jet}=\xi L_{jet}$,
where $\zeta$ is the acceleration efficiency and $\xi=\zeta c_{pp}$ is the
efficiency of \gray production. Assuming $c_{pp} \approx \zeta \approx 10\%$
($\xi \approx 10^{-2}$), and a peak \gray isotropic luminosity above 100
MeV of $L_{\gamma} \thicksim 10^{36}~\rm{erg\:s^{-1}}$, the corresponding jet power is
$L_{jet} \thicksim 10^{38}~\rm{erg\:s^{-1}}$, which is an order of magnitude lower
than the Eddington accretion limit for the system (assuming that the compact object
in \cyg is a black hole with a mass of $M_x = 10~M_{\odot}$, where $M_{\odot}$ is the solar mass)
and it is consistent with the average bolometric luminosity of the
hypersoft state, $L^{HYS}_{bol} \approx 1.2 \times 10^{38}~\rm{erg~s^{-1}}$
\citep{koljonen_10}. Furthermore, as demonstrated by \citet{cerutti_11},
it is unlikely that the HE \grays have a coronal origin (unless the corona is unrealistically
extended): the observed \gray emission is linked to the physics of the jet and created outside
the \gray photosphere (at distances greater than $10^8$-$10^{10}$ cm from the compact object).

\lat observations of the source revealed a \gray orbital
modulation \citep{abdo_09} coherent with the orbital period ($t_{mod}=4.8$
hr).
The folded emission above 100 MeV is characterized by a sharp maximum in correspondence
to the superior conjunction of the system (compact object behind the WR star, with respect to the line of sight).
The \grays appear to be approximately in antiphase with
the X-ray modulation, a fact that may be linked to the
different physical origin of the two components. The modulated
\grays are produced only if the protons are confined in the binary
system in time scales less than $\xi\: t_{mod}$ otherwise the
protons will escape from the binary system.
Since the cooling time for $pp$ interactions is $t_{pp}\approx10^{15}/N\:\:\rm s$ ($N=n_{H}/1\:\rm cm^{-3}$), from the relation $t_{pp}=\xi\:t_{mod}$,
the condition for \gray modulation is satisfied only for densities ($n_H$) larger than $\approx6\times10^{12}$
$\rm{cm}^{-3}$ (assuming $\xi\approx10^{-2}$ and $t_{mod}=4.8 \rm h$).
Accordingly, this condition is satisfied especially near superior conjunction
because of an increase of the density along the orbit
(anisotropic wind) or due to  protons interacting with
clumps (regions where the density is significantly higher than
average value) \citep{ar09}. At other phases along the orbit, the density of the wind
should be significantly lower and since this change of density causes the change of luminosity of produced \grays,
this results a modulation of \gray signal. 

\section{Production and absorption of gamma-rays.} \label{sec3}

In the binary systems the production of \grays (both in leptonic
and hadronic origin) is accompanied by strong absorption of these
photons. Depending on both their energy and the site of HE
production, the emitted \grays can be absorbed by interactions
with the X-ray photons from the corona/disk complex or with the UV
photon field from the companion star. We discuss below the
production of \grays from $pp$ interaction as well as the opacity
of photon absorption.

\subsection{Production of gamma-rays and neutrinos from $pp$ interaction}
Hadronic inelastic scattering, between high-energy protons
(accelerated in the jet) with cold protons from the WR wind,
is responsible for the production of secondary \grays and
neutrinos. The fluxes of produced particles are calculated using
the analytical approximation derived in \citep{kelner_06}, obtained
from numerical simulations of $pp$ interaction with the publicly
available code SIBYLL. The analytical formulae provide a very
good description of the flux and energy distribution of secondaries
for energies above 100 GeV. The formula for \grays also includes the
contribution of $\eta$ meson decay, in addition to that of $\pi^0$,
with an overall accuracy of order a few percent.\\
At energies below 0.1 TeV and down to the rest energy of the
$\pi$-meson, the fluxes of \grays and neutrinos are modeled with delta function
approximation as suggested by \citep{kelner_06}, namely the fluxes are given by:
\begin{equation}
\Phi_{i} (E_i) =\frac{c\,A_{i}\:\kappa\:n_{H}}{4\pi\:D^2\:K_{\pi}}\int_{E_{\pi,\rm th}}^\infty
\frac{\sigma_{pp}(E_c)\:N_{p}(E_c)}
{\sqrt{E_\pi^2 - m_\pi^2c^4}}\de E_\pi\,
\label{gamma-delta}
\end{equation}
for both particle type $i$ ($i=\gamma,\nu$), $E_{\pi,\rm th}=
\frac{E_{i}}{1-r_{i}}+(1-r_{i})\frac{m_\pi^2c^4}{4\:E_{i}}$
where $r_{\gamma}=0$, $r_{\nu}=(m_{\mu}/m_{\pi})^{2}$, $A_{\gamma}=2$,
and $A_{\nu}=(1-r_{\nu})^{-1}$. In Eq.~(\ref{gamma-delta}) $K_{\pi}$ is the
mean fraction of proton kinetic energy transferred to pions, $c$ is the speed of light in a vacuum, $D$ is the distance from the source, $\kappa$ is the free parameter to match the results of Monte-Carlo simulations \citep{kelner_06},
$\sigma_{pp}$ is the $pp$ inelastic interaction cross section,
$E_{c}=m_{p}c^2+E_{\pi}/K_{\pi}$ and $m_p,\:m_{\mu},\:m_{\pi}$
are the proton, muon and pion masses, respectively.
\subsection{Absorption of gamma-rays}

The produced \grays can be absorbed by interactions with UV stellar photons
from the companion star. In the calculations below, we take into account this
absorption by using the opacity averaged over the injection
angles:
\begin{equation}
\tau_\gamma(E_\gamma,r)=\int_r^\infty \int_{\epsilon_{min}}^\infty
n(\epsilon_0,\,r') \sigma_{\gamma\gamma}(\epsilon_0,E_\gamma) \de\epsilon_0\: \de r',
\label{tau}
\end{equation}
where $\epsilon_0$ is energy of the companion star's
photons, $\epsilon_{min}$ is the threshold of pair production,
$\epsilon_{min}=m_{e}^2c^4/E_{\gamma}$, $E_\gamma$ is the energy
of the \gray, $m_ec^2$ is the electron rest energy and $\sigma_{\gamma\gamma}(\epsilon_0,E_\gamma)$ is
the cross section for photon-photon pair production
\citep{Gould67}. The distribution of stellar photons is assumed to
have a blackbody spectrum peaking at the star's effective
temperature ($T_{\rm eff}$):
\begin{equation}
n(\epsilon_0,r)=\frac{2 \pi
\epsilon_0^{2}}{(h\:c)^3}\frac{1}{e^{\epsilon_0/k\:T_{\rm eff}}-1}\frac{R^2_\star}{r^2},
\end{equation}
where $h$ and $k$ are the Planck and Boltzmann constants respectively,
and $R_\star$ is the radius of the companion star. For \cyg we adopt the following
values: $T_{\rm eff}=10^5~\rm{K}$, and $R_\star=6\times10^{10}$
cm.
\begin{figure}[thb]
   \centering
    \includegraphics[width=130mm,angle=0]{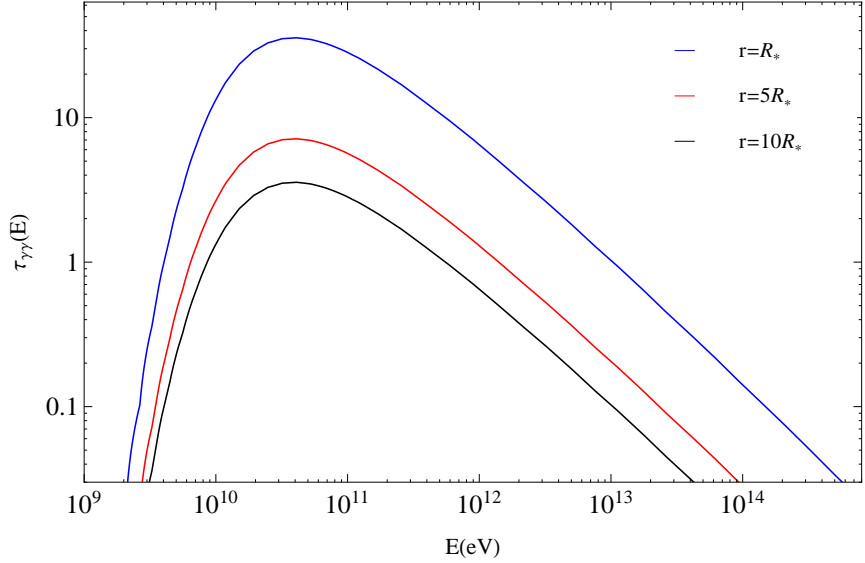}
   \caption{The opacity of photon-photon pair production
   (averaged over injection angles) calculated for different distance from the companion star
   for the \cyg system}.
    \label{abs}
\end{figure}

This absorption depends strongly on the geometry.
It depends on the relative location of the \gray source, the
companion star, and the line of sight to the observer.
We calculated the opacity, which depends on the distance from the
companion star ($r$), by averaging over the injection
angles. This is illustrated in Fig.~\ref{abs} where the opacity is
calculated from Eq.~\ref{tau} for $r=R_\star$, $5R_\star$ and
$10R_\star$. The VHE \grays produced very close to star (e.g.
$r=R_\star$) will be heavily attenuated
($\tau_{\gamma\gamma}\approx30$ for $E_\gamma=100\:\rm{GeV}$), and
the observed spectrum of VHE photons would be hardened compared to
its intrinsic shape. However this opacity will vary depending on
the distance, for comparison, the opacity of gamma-photon
interaction at distance $10R_\star$ is $\tau_{\gamma\gamma}\approx
3.0$ (at 100 GeV). It rapidly drops for higher energies, and the
source becomes transparent for \grays above 10 TeV.
It should be noted that a similar study of pair-production opacity
along the orbit presented by \citet{zdziarski_12} found consistent results:
the opacity peaks around 100 GeV and drops at higher energies.

\section{The results}\label{sec4}

In our model we assume that the energy distribution of accelerated protons
follows a power law plus a high-energy cut-off at 100 TeV:
\begin{equation}
N_{p}(E_p) \thicksim E_{p}^{-\alpha} \exp\left(-\frac{E_{p}}{100 \:\rm{TeV}}\right).
\label{fin}
\end{equation}
Using the proton distribution given by Eq.~(\ref{fin}), the
corresponding fluxes of HE \gray and neutrinos are calculated. In
our calculations, we focus on $\nu_{\mu}$ and ${\bar \nu}_\mu$ for
which neutrino detectors are most sensitive.

\begin{figure}[thb]
   \centering
    \includegraphics[width=130mm,angle=0]{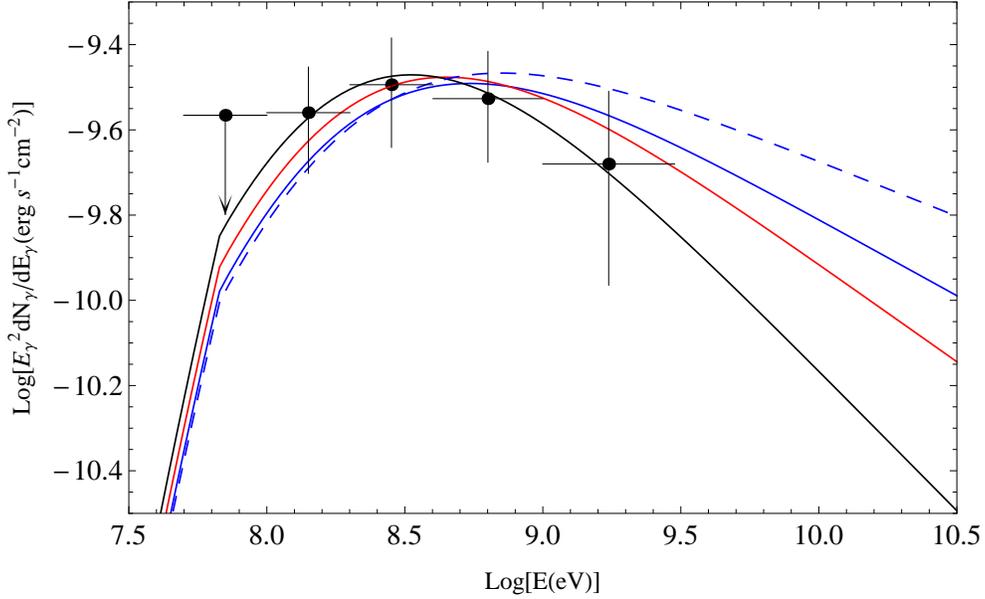}
   \caption{Hadronic modeling of the \gray flaring
   spectrum detected by AGILE assuming different indices of the
   initial proton energy distribution. The blue, red and black solid lines
   correspond to power-law indices $\alpha=2.4$, $2.5$ and $2.7$,
   respectively;   the dashed line corresponds to $\alpha=2.3$.}
    \label{pl}
\end{figure}

As discussed in the previous section, \grays produced from $pp$
interaction will be effectively absorbed by stellar photon field
and this absorption amplitude depends on the distance from the
star where these photons are created. Accordingly, below we
discuss different possibilities for the \gray absorption in
\cyg, assuming that gamma-photon interactions occur close
(strong absorption) or far away from the companion star (weak absorption).\\
At the same time, since the cross section of $\gamma\gamma$
interaction achieves its maximum at $\epsilon
E/(m_ec^{2})^2\approx4$, the absorption of GeV \grays will be
effectively in the X-ray photon field. However, since in our model
the \grays are produced farther from the compact object, where the
density of X-ray photon field is low, these photons escape the
region without significant absorbtion. Accordingly, using the
AGILE observations of \cyg \gray flares, the limit on
the power-law index of proton energy distribution can be derived.
The resultant fits of the data for different initial proton energy
distributions are shown in Fig.~\ref{pl}.
As one can see, the minimum power-law index which can reproduce
the observed data corresponds to $\alpha=2.4$ (blue solid line),
since for lower power-law indices predicted flux of \grays
is larger than the \gray data (blue dashed line in Fig.~\ref{pl}).
Moreover, softer proton spectra fit better the data (see the red and black
lines in Fig.~\ref{pl}, calculated for $\alpha$ = 2.5 and 2.7 respectively).

\begin{figure*}[thb]
   \centering
   \subfigure[]{\includegraphics[width=80mm,angle=0]{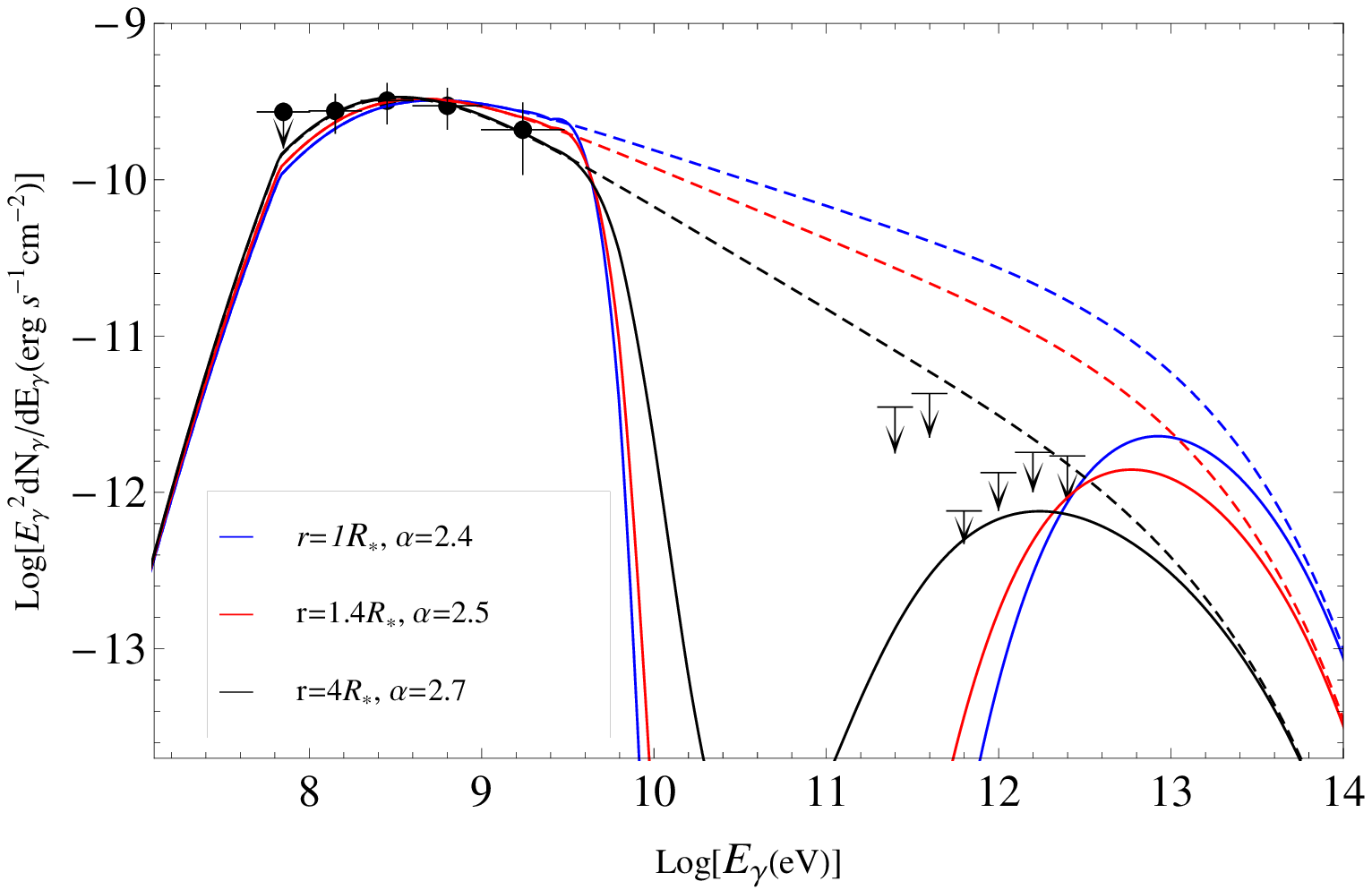}}
   \subfigure[]{\includegraphics[width=80mm,angle=0]{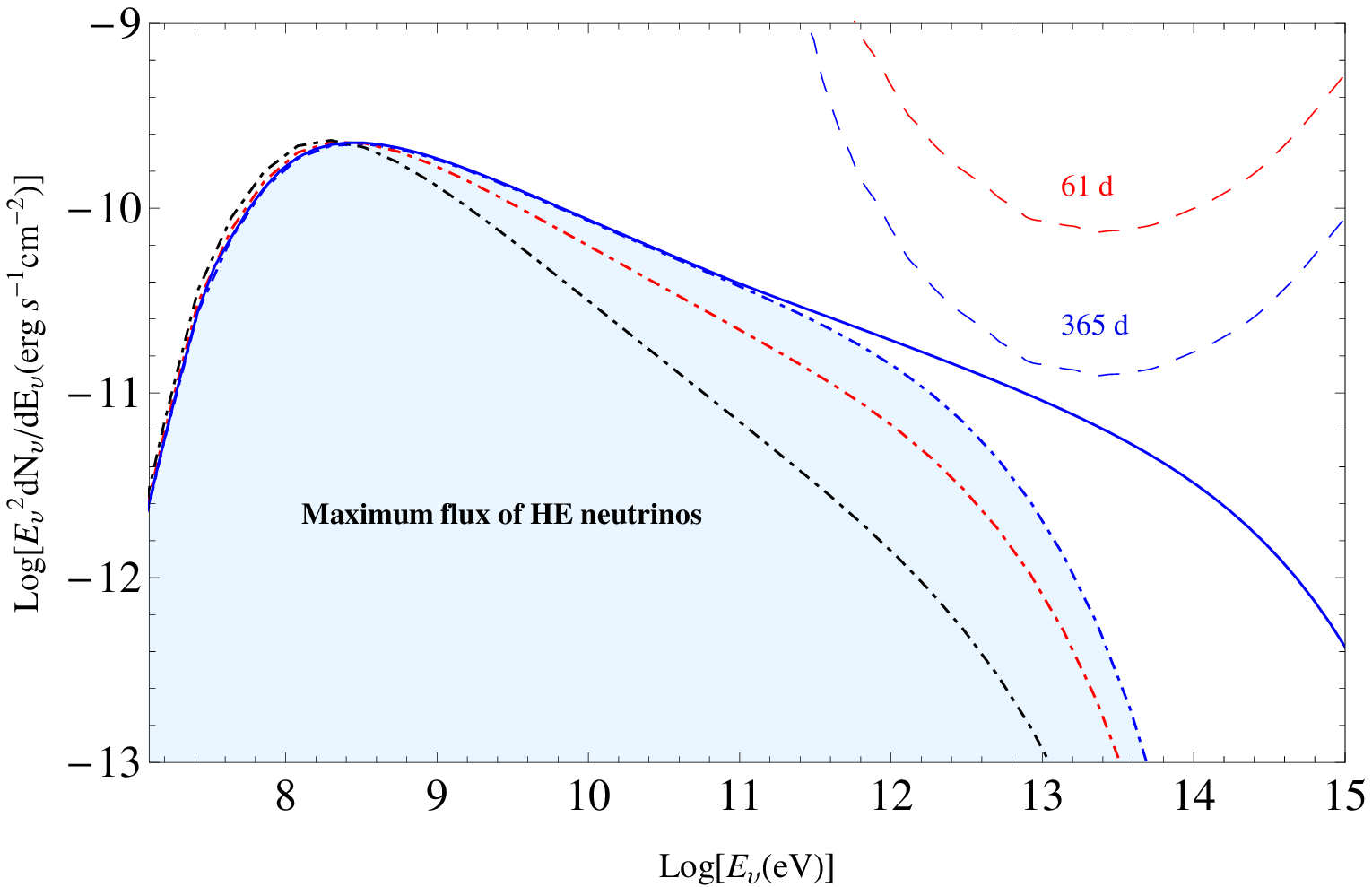}}

   \caption{Panel (a): Calculated \gray unabsorbed (dashed lines) and absorbed
   (solid lines) emission shown with the \gray flaring flux ($\sim$8 days)
   detected by  AGILE  \citep{piano_12}, and the MAGIC upper limits
   obtained when \cyg was in the soft state (four observation cycles,
   corresponding to a total of 30.8 hours, \citealp{aleksic_10}).
   The \gray absorption by UV stellar photons is calculated for different
   distances $r$ from the companion star. In panel (b)
   the corresponding calculated spectrum of HE neutrinos is presented,
   the filled area represents the maximum flux of HE neutrinos.
   In panel (b) the solid blue line corresponds to a neutrino model
   with the same parameters of the blue dot-dashed curve
   ($r=R_\star$, $\alpha=2.4$), but with a maximum energy of protons of $10\:\rm{PeV}$. Red and blue dashed lines in panel (b) corresponds to IceCube sensitivities expected for two-month (61 days) and one-year (365 days) exposure time, respectively.}
   \label{fg}
\end{figure*}

Fig.~\ref{fg}~(a) shows the \gray fluxes calculated for the proton
energy distributions given by Eq.~\ref{fin} with $\alpha=2.4$, $2.5$ and $2.7$
(blue, red and black colors respectively). The dashed lines corresponds to the
unabsorbed flux of \grays, instead the corresponding absorbed spectra
are depicted with solid lines. As one can see, in all cases the predicted
unabsorbed flux of \grays at TeV energies is larger than the
ULs derived from MAGIC observations. However, taking into account
the absorption of \grays, our model is in agreement with
the observed data in MeV/GeV and TeV energies. For example, in the case of a
proton index of $\alpha=2.4$ (which is related to the highest \gray flux),
assuming that the \grays are produced at the distance $r = R_\star$,
the density of stellar photon field is so high that the absorbed \gray spectrum is lower than the
MAGIC ULs (blue solid line in Fig.~\ref{fg}~(a)). Similar results
are obtained for values of $\alpha=2.5$ with $r = 1.4 R_\star$ (red
line in Fig.~\ref{fg}~(a)), and for $\alpha=2.7$ with $r = 4
R_\star$ (blue line in Fig.~\ref{fg}~(a)). We remark that the
values presented here corresponds to the maximum distances from
the star where the \grays can be created:
for closer distances, the absorption is higher and the predicted flux will be smaller.\\
Neutrinos are produced together with the \grays, but unlike
the \grays they escape from the region without any absorption. In
Fig.~\ref{fg}~(b) we show the resulting neutrino fluxes
from \cyg corresponding to the \gray models of Fig.~\ref{fg}~(a).
Since the minimum power-law index obtained from the \gray
observations corresponds to $\alpha=2.4$, the predicted flux of
neutrinos can be considered as a maximum flux during the \gray activity of
the microquasar (this is shown with a filled area in Fig.~\ref{fg}~(b)).
This predicted flux of HE neutrinos is compared with the IceCube sensitivities
expected for different exposure times: two-month (61 days)
and one-year (365 days) exposure time (red and blue dashed
lines in Fig.~\ref{fg}~(b), respectively). The effective area for
the 40-string configuration \citep{abbas_11} have been scaled to
the 86-string configuration\footnote{The real
effective area related to the 86-string  IceCube configuration for
a point source may be marginally different, but so far no public
data are available.} (full string configuration).
The maximum predicted flux of neutrinos is almost two order of magnitude
less than the 61 day IceCube sensitivity, in agreement with the absence of a detectable
neutrino signal from \cyg from  the current IceCube observations
\citep{abbas_13}. Nevertheless, we can assume that relativistic
particles inside the jet can be accelerated far above 100 TeV,
reaching PeV energies (supposing that \cyg is a Galactic
``Pevatron''). Accordingly, if the cut-off energy in the proton
spectrum (see Eq.~\ref{fin}) is at 10 PeV, the predicted flux of
HE neutrinos is slightly lower than the IceCube sensitivity for
one-year exposure time (solid blue line in Fig.~\ref{fg}~(b)).
Therefore, future detection of HE neutrinos from \cyg is possible
if the particles in the jet are accelerated up to ultra high energies.

In the previous discussion only the \gray data from AGILE
observations are used. However, the derived conclusions are valid
also for the \gray spectrum obtained by \lat during a
prolonged \gray activity of the microquasar \citep{abdo_09}.
The predicted flux of HE neutrinos, related to the \gray activity detected by
\lat, will be slightly lower than the ones presented in
Fig.~\ref{fg}~(b), since the  \lat photon index for the
average emission from \cyg above 100 MeV is $\Gamma=2.7$
\citep{abdo_09}, instead of the value $\Gamma = 2.0$
determined  by AGILE in the range 100 MeV - 1 GeV.

\section{Discussion}\label{sec5}

\begin{figure*}[t]
\begin{center}
 \includegraphics[width=16.5cm]{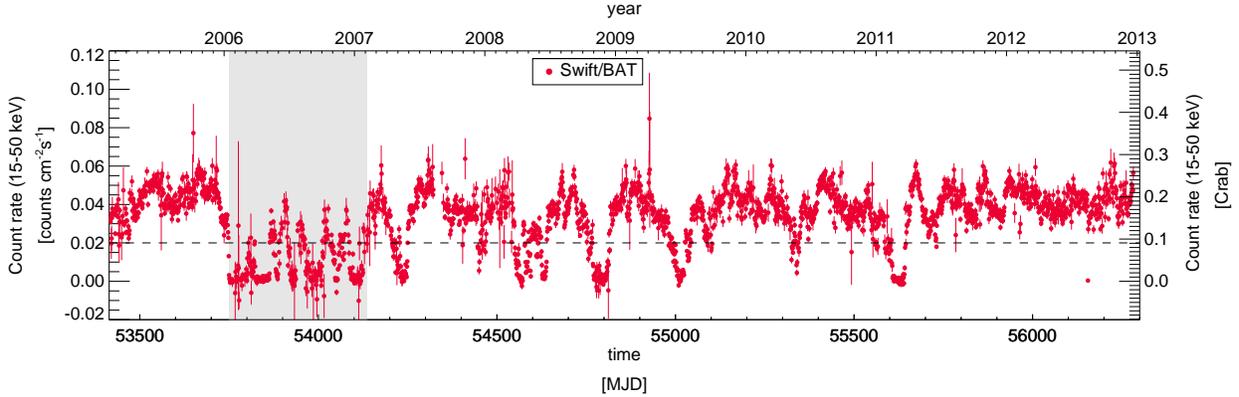}
 \caption{Hard X-ray light curve of \cyg as detected by
 \textit{Swift}/BAT (15-50 keV) between February 2005 and January 2013.
 The horizontal dashed line represents the transition level
 ($0.02~\mathrm{counts~cm^{-2}~s^{-1}} \approx0.091~\mathrm{Crab}$) between soft and hard X-ray spectral state.
 Gamma-ray emission above 100 MeV is expected during minima of the hard X-ray curve \citep{piano_12}.}
 \label{cyg_x3_bat}
\end{center}
\end{figure*}

Hadronic \gray emission from \cyg, discussed and presented in the
previous sections, requires an effective acceleration of hadrons (protons) in the jet
of the microquasar. The total energy of the protons (for
$\alpha=2.4$) corresponds to $W_{p}\approx 1.74\times10^{40}$ erg,
assuming the number density of the wind to be $n_{H}=6\times10^{12}$
$\rm{cm}^{-3}$.
The kinetic power of protons in the jet ($L_{p} = W_{pp}/t_{pp}$)
would be $L_{p}=1.04\times10^{38}\:\rm{erg\:s^{-1}}$, consistent with
the hypersoft state bolometric luminosity.
This flux of \grays would be accompanied by the flux of HE neutrinos and, for a proton injection
power $L_{p}=1.04\times10^{38}\:\rm{erg\:s^{-1}}$, the predicted
neutrino flux is $f_{10\:TeV}\approx9.1\times10^{-12}\:\rm{erg\:cm^{-2}\:s^{-1}}$
at 10 TeV (solid line in Fig.~\ref{fg}~(b)). 
Furthermore, the minimum detectable neutrino flux for a neutrino point source with a generic $E^{-2}$ spectrum, after one year of operation, is $f_{sens}=2.72\times10^{-11}\:\rm{erg\:cm^{-2}\:s^{-1}}$ at 5 $\sigma$ significance \citep{ahrens}.
We then obtain $f_{10\:\rm{TeV}}/f_{sens}\approx 0.3$. This estimate shows
that only a quite hard proton injection rate (of index $\alpha <
2.4$) extending up to 10 PeV can produce a detectable flux of HE neutrinos from \cyg. The
flux of HE neutrinos depends, not only on the proton injection
rate (which may be time variable), but also on the duration
of the HE activity of the source. It is known that the microquasar
\cyg emits \grays only in specific X-ray conditions: namely,
during bright soft X-ray spectral states coincident with
\textit{minima} of the hard X-ray light curve
\citep{tavani_09,abdo_09,corbel_12,piano_12}. In principle, these
states could last several months \citep{abdo_09}
or even longer, with a strong probability of emitting \grays. These
prolonged episodes of minimal hard X-ray emission (and corresponding maximal
soft X-ray emission) might imply a significant increase of neutrino emission which can be detected by
IceCube under favorable conditions. Interestingly, the
hard X-ray light curve as detected by \textit{Swift}/BAT (15-50
keV) (see Fig.~\ref{cyg_x3_bat}) shows such a prolonged activity.
Namely, with the gray region of the plot is indicated a prolonged
period (MJD: $\sim$53749--54136, between 2006 and February 2007) in which \cyg
is found to be most of the time ($\sim$70\%) in a soft state
(\textit{Swift}/BAT count rate $\lesssim~0.02~\mathrm{counts}$ $\mathrm{cm^{-2}~s^{-1}}$).
Therefore, if the conditions for \gray emission discussed by
\citealp{piano_12} are valid, we deduce that in that period \cyg was characterized by a
quasi-continuous emission of \grays and possibly detectable
neutrinos. Unfortunately neither AGILE nor \fermi (nor
sensitive neutrino detectors) were operational in 2006 to test this picture.
In the future, such a possible prolonged active state
accompanied by a quasi-continuous emission of \grays (of
hadronic origin) might reach a neutrino flux close to
detection by instruments such as IceCube.

\newpage

\section{Conclusions}\label{sec6}

Effective particle acceleration in microquasar jet makes
these objects strong sources of MeV-TeV \grays. These \grays can be
produced via leptonc interactions (e.g. IC scattering of low energy photons
by relativistic electrons), as well as via hadronic processes
(e.g., $p\gamma$ or $pp$ interactions). In case of hadronic interactions,
the flux of \grays is accompanied by emission of HE neutrinos.
Therefore, these sources are interesting target of observations with HE neutrino detectors.

We investigated in this paper the possibility of detecting
HE neutrinos from \cyg within a hadronic model of emission.
We discussed a simplified picture of hadronic \gray production,
namely we assumed that the protons are effectively accelerated by
the jet up to 100 TeV energies (the achieved maximum
energy of protons depends on the magnetic field and on the size of jet).
These protons can escape from the jet and interact with cold protons in the wind of the companion star.
From these inelastic collisions, neutral and charged pions are produced
with subsequent substantial neutrino emission.
In this scenario, the effective production of \grays occurs
only if the surrounding matter (number) density is larger than
$\approx6\times10^{12}$ $\rm{cm}^{-3}$ (corresponding to a cooling time
$\leq\xi\:t_{mod}$, with $\xi \approx 10^{-2}$).
This condition is expected to be satisfied for \cyg at superior conjunction
where the modulated \gray emission along the orbit reaches its maximum \citep{abdo_09}.
The absorption of \grays (by UV stellar photons from the WR star) does
not affect the propagation of photons at MeV/GeV energies (unlike
the TeV \grays), therefore the minimum index of the initial proton
energy distribution can be derived from the \cyg \gray flares to be
$\alpha\geq2.4$ (from AGILE data). The spectrum of \grays and high-energy neutrinos
can be then calculated. Taking into account the absorption of TeV \grays
as deduced from the MAGIC upper limits, we can constrain the distance from the companion star
where the \grays should be created. Within this distance, the
absorption modifies the spectrum at TeV energies. The injection
rate of protons should be $L_p\approx10^{38}\:\rm{erg\:s^{-1}}$ in
order to explain the observed spectrum of \grays. The required
power is physically reasonable: it is consistent with the
bolometric luminosity during the hypersoft spectral state
(correlated to the \gray transient activity, \citealp{piano_12}),
and it is lower than the Eddington accretion limit for a stellar
black hole mass in \cyg.
Together with \grays, HE neutrinos are produced, escaping the region
without any absorption. We found that a maximal neutrino flux
(expected during the \gray activity of \cyg) corresponds to an
accelerated proton distribution with $\alpha=2.4$. In case
of short exposure time (two months), the predicted flux of
HE neutrinos is almost two orders of magnitude less than
the 61-day IceCube sensitivity. Only assuming that protons are accelerated up to
10 PeV energies with a spectrum harder than $\alpha = 2.4$,
the predicted neutrino flux would be detectable by
the IceCube full-string sensitivity (with 1-year  exposure time).
Whether a proton spectrum harder than $\alpha = 2.4$ can be
produced in \cyg in a time variable fashion not to contradict TeV
upper limits is an open question that will be investigated by
future observations.
Long term observations of \cyg with IceCube combined with GeV and TeV observations can give
important information about emission of neutrinos from microquasars, providing invaluable
constraints on the hadronic particle density in relativistic jets.
These considerations show how \cyg is a crucially interesting source, not only for
radio-to-\gray observations, but also for new-generation neutrino detectors.

\section*{Acknowledgements}
We warmly thank Felix Aharonian for collaboration and extensive discussions on
the physics of the \cyg system.\\
Research partially supported by the ASI grants n. I/042/10/0 and I/028/12/0.


\end{document}